\input{aipcheck}

\documentclass[
    ,final            
  ]
  {aipproc}

\layoutstyle{6x9}

\begin{document}

\title{Update of $\eta$-$\eta^\prime$ mixing from $J/\psi\to VP$ decays\footnote{
Talk given at SCADRON 70: Workshop on "Scalar Mesons and Related Topics", 
Lisbon, Portugal, 11-16 February 2008}}


\author{R.~Escribano}{
  address={Grup de F\'{\i}sica Te\`orica and IFAE, Universitat Aut\`onoma de Barcelona,\\
                    E-08193 Bellaterra (Barcelona), Spain}
}

\begin{abstract}
The $\eta$-$\eta^\prime$ mixing angle is deduced from an updated phenomenological analysis
of $J/\psi$ decays into a vector and a pseudoscalar meson.
Corrections due to non-ideal $\omega$-$\phi$ mixing are confirmed to be crucial to find
$\theta_P=(-16.9\pm 1.7)^\circ$, in agreement with most recent analyses.
The newly reported values of $B(J/\psi\to\rho\pi)$ by the BABAR and BES Collaborations make more difficult a reasonable description of data.
\end{abstract}

\maketitle


\section{Introduction}
The value of the $\eta$-$\eta^\prime$ mixing angle $\theta_P$ in the pseudoscalar-meson nonet has been discussed many times in the last forty years.
A well-known contribution to this discussion is the phenomenological analysis performed by Gilman and Kauffman \cite{Gilman:1987ax} two decades ago.
The approximate value $\theta_P\simeq -20^\circ$ was proposed by these authors through a rather
exhaustive discussion of the experimental evidence available at that time.
Another analysis by Bramon and Scadron \cite{Bramon:1989kk} concluded that a somewhat less negative value, $\theta_P=-14^\circ\pm 2^\circ$, seems to be favoured.
A significant difference between these two independent analyses concerns the set of rich data on
$J/\psi$ decays into a vector and a pseudoscalar meson, $J/\psi\to VP$, which were included in the first analysis \cite{Gilman:1987ax} but not in the second one \cite{Bramon:1989kk}.
Ten years ago, these authors together with the present author deduced the value of the $\eta$-$\eta^\prime$ mixing angle from this relevant set of $J/\psi\to VP$ decay data including for the first time corrections due to non-ideal $\omega$-$\phi$ mixing \cite{Bramon:1997mf}.
These corrections turned out to be crucial to find $\theta_P=(-16.9\pm 1.7)^\circ$, 
which was appreciably less negative than previous results coming from similar analyses.
The purpose of the present contribution is to update the phenomenological analysis done in 1997
of $J/\psi\to VP$ decays, with $V=\rho, K^\ast, \omega, \phi$ and $P=\pi, K, \eta, \eta^\prime$,
aimed at determining the quark content of the $\eta$ and $\eta^\prime$ wave functions.
In this sense, we will follow quite closely the analysis in Ref.~\cite{Bramon:1997mf} except that now we will use the most recent experimental data accounting for these decays \cite{Yao:2006px}.

\section{Notation}
\label{notation}
We work in the quark-flavour basis consisting of the states
\begin{equation}
\label{mathstates}
|\eta_q\rangle\equiv\frac{1}{\sqrt{2}}|u\bar u+d\bar d\rangle\ ,\qquad
|\eta_s\rangle\equiv |s\bar s\rangle\ .
\end{equation}
The physical states $\eta$ and $\eta^\prime$ are assumed to be the linear combinations
\begin{equation}
\label{physicalstates}
|\eta\rangle=X_\eta |\eta_q\rangle+Y_\eta |\eta_s\rangle\ ,\qquad
|\eta^\prime\rangle=X_{\eta^\prime}|\eta_q\rangle+Y_{\eta^\prime}|\eta_s\rangle\ ,
\end{equation}
with $X_{\eta (\eta^\prime)}^2+Y_{\eta (\eta^\prime)}^2=1$.
The implicit assumptions in Eq.~(\ref{physicalstates}) are the following:
i) absence of gluonium in the $\eta$ and $\eta^\prime$ wave functions,
ii) no mixing with $\pi^0$ ---isospin symmetry, and
iii) no mixing with radial excitations or $\eta_c$ states.
In absence of gluonium ---standard picture---
the coefficients $X_{\eta (\eta^\prime)}$ and $Y_{\eta (\eta^\prime)}$ are described in terms of a single angle,
$X_\eta=Y_{\eta^\prime}\equiv\cos\phi_P$ and $X_{\eta^\prime}=-Y_\eta\equiv\sin\phi_P$, thus
\begin{equation}
\label{physicalstatesphiP}
|\eta\rangle=\cos\phi_P |\eta_q\rangle-\sin\phi_P |\eta_s\rangle\ ,\qquad
|\eta^\prime\rangle=\sin\phi_P |\eta_q\rangle+\cos\phi_P |\eta_s\rangle\ ,
\end{equation}
where $\phi_P$ is the $\eta$-$\eta^\prime$ mixing angle in the quark-flavour basis.
It is related to its octet-singlet analog $\theta_P$ through
\begin{equation}
\label{angle}
\theta_P=\phi_P-\arctan\sqrt{2}\simeq\phi_P-54.7^\circ\ . 
\end{equation}
Similarly, for the vector states $\omega$ and $\phi$ the mixing is given by
\begin{equation}
\label{physicalstatesphiV}
|\omega\rangle=\cos\phi_V |\omega_q\rangle-\sin\phi_V |\phi_s\rangle\ ,\qquad
|\phi\rangle=\sin\phi_V |\omega_q\rangle+\cos\phi_V |\phi_s\rangle\ ,
\end{equation}
where $\omega_q$ and $\phi_s$ are the analog non-strange and strange states of
$\eta_q$ and $\eta_s$, respectively.

\section{Experimental data}
We use the most recent experimental data available for $J/\psi\to VP$ decays taken from
Ref.~\cite{Yao:2006px}.
The data for the $K^{\ast +}K^-+\mbox{c.c.}$, $K^{\ast 0}\bar K^0+\mbox{c.c.}$,
$\rho\eta$ and $\rho\eta^\prime$ channels remain the same since 1996 \cite{Barnett:1996hr}
and were reported by the DM2 \cite{Jousset:1988ni} and Mark III \cite{Coffman:1988ve} Collaborations.
The new measurements come from the BES Collab.,
Ref.~\cite{Ablikim:2005pr} for $\omega\pi^0$, $\omega\eta$ and $\omega\eta^\prime$ and
Ref.~\cite{Ablikim:2004hz} for $\phi\eta$, $\phi\eta^\prime$ and the upper limit of $\phi\pi^0$,
and the BABAR Collab.~for $\omega\eta$ \cite{Aubert:2006jq}.
The BES data are based on direct $e^+e^-$ measurements,
$e^+e^-\to J/\psi\to\omega\pi^0, \omega\eta, \omega\eta^\prime$ for channels involving $\omega$, 
$e^+e^-\to J/\psi\to\mbox{hadrons}$ for $\phi$, and
the $e^+e^-\to J/\psi\to\phi\gamma\gamma$ for the upper limit of $\phi\pi^0$.
The BABAR data is obtained with the initial state radiation method to lower the center-of-mass energy to the production ($J/\psi$) threshold.
Special attention is devoted to the case of $\rho\pi$.
Four new contributions have been reported since the old weighted average
$B(J/\psi\to\rho\pi)=(1.28\pm 0.10)\%$ \cite{Barnett:1996hr},
$(2.18\pm 0.19)\%$ from $e^+e^-\to\pi^+\pi^-\pi^0\gamma$ \cite{Aubert:2004kj},
$(2.184\pm 0.005\pm 0.201)\%$ from $e^+e^-\to J/\psi\to\pi^+\pi^-\pi^0$ \cite{Bai:2004jn},
$(2.091\pm 0.021\pm 0.116)\%$ from $\psi(2S)\to\pi^+\pi^-J/\psi$ \cite{Bai:2004jn}
---the weighted mean of these two measurements is $(2.10\pm 0.12)\%$, and
$(1.21\pm 0.20)\%$ from $e^+e^-\to\rho\pi$ \cite{Bai:1996rd}.
Thus, the new weighted average is $(1.69\pm 0.15)\%$ with a confidence level of 0.001
\cite{Yao:2006px}.
In Table \ref{tableB}, we show the branching ratios for the different decay channels
according to the present-day values \cite{Yao:2006px} (third column)
and the old ones \cite{Barnett:1996hr} (fourth column).
\begin{table}
\begin{tabular}{cccc}
\hline
\tablehead{1}{c}{b}{Channel} &
\tablehead{1}{c}{b}{Fit} &
\tablehead{1}{c}{b}{Exp.~2008 \cite{Yao:2006px}} &
\tablehead{1}{c}{b}{Exp.~1997 \cite{Barnett:1996hr}} \\
\hline
$\rho\pi$ 						& $12.7\pm 1.2$		
							& $16.9\pm 1.5$ 			& $12.8\pm 1.0$ \\
$K^{\ast +}K^-+\mbox{c.c.}$		& $5.4\pm 0.7$			
							& $5.0\pm 0.4$				& $5.0\pm 0.4$ \\
$K^{\ast 0}\bar K^0+\mbox{c.c.}$ 	& $4.8\pm 0.7$			
							& $4.2\pm 0.4$				& $4.2\pm 0.4$ \\
$\omega\eta$ 					& $1.76\pm 0.18$		
							& $1.74\pm 0.20$			& $1.58\pm 0.16$ \\
$\omega\eta^\prime$ 			& $0.187\pm 0.055$		
							& $0.182\pm 0.021$			& $0.167\pm 0.025$ \\
$\phi\eta$ 					& $0.68\pm 0.16$		
							& $0.74\pm 0.08$			& $0.65\pm 0.07$ \\
$\phi\eta^\prime$ 				& $0.29\pm 0.11$		
							& $0.40\pm 0.07$			& $0.33\pm 0.04$ \\
$\rho\eta$ 					& $0.219\pm 0.019$ 	
							& $0.193\pm 0.023$			& $0.193\pm 0.023$ \\
$\rho\eta^\prime$ 				& $0.102\pm 0.011$ 	
							& $0.105\pm 0.018$			& $0.105\pm 0.018$ \\
$\omega\pi^0$ 					& $0.39\pm 0.03$		
							& $0.45\pm 0.05$			& $0.42\pm 0.06$ \\
$\phi\pi^0$ 					& $0.0010\pm 0.0001$	
							& $<0.0064$\ C.L.~90\%		& $<0.0068$\ C.L.~90\% \\
\hline
$g$ 			& $1.148\pm 0.053$ 	&	$s$			&	$0.141\pm 0.038$ \\
$e$ 			& $0.120\pm 0.004$ 	&	$r$			&	$-0.164\pm 0.014$ \\
$\theta_e$ 	& $1.36\pm 0.15$ 		&	$\theta_P$	&	$(-16.9\pm 1.7)^\circ$ \\
\hline
\end{tabular}
\caption{
Experimental $J/\psi\to VP$ branching ratios (in units of $10^{-3}$) from
Ref.~\cite{Yao:2006px} (third column) and Ref.~\cite{Barnett:1996hr} (fourth column).
The results of our best fit (second column), with fixed values for
$x=0.81\pm 0.05$ and $\phi_V=(3.2\pm 0.1)^\circ$,
and the fitted values of the parameters (lower part) are also shown.}
\label{tableB}
\end{table}

\section{Phenomenological model}
Since the $J/\psi$ meson is an almost pure $c\bar c$ state, its decays into $V$ and $P$ are 
Okubo-Zweig-Iziuka (OZI) rule-suppressed and proceed through a three-gluon annihilation diagram and an electromagnetic interaction diagram.
Aside from the OZI-suppressed diagrams common to all hadronic $J/\psi$ decays,
the doubly disconnected diagram,
where the vector and the pseudoscalar exchange an extra gluon,
is also expected to contribute to the $J/\psi$ decays\footnote{
We will not consider here the doubly disconnected diagram representing the diagram connected to a possible glueball state.}.
The amplitudes for the $J/\psi\to VP$ decays are expressed in terms of an
$SU(3)$-symmetric coupling strength $g$ (SOZI amplitude)
which comes from the three-gluon diagram,
an electromagnetic coupling strength $e$ (with phase $\theta_e$ relative to $g$)
which comes from the electromagnetic interaction diagram \cite{Kowalski:1976mc}, and
an $SU(3)$-symmetric coupling strength which is written by $g$ with suppression factor $r$ contributed from the doubly disconnected diagram (nonet-symmetry-breaking DOZI amplitude) 
\cite{Haber:1985cv}.
The $SU(3)$ violation is accounted for by a factor $(1-s)$ for every strange quark contributing to $g$,
a factor $(1-s_p)$ for a strange pseudoscalar contributing to $r$,
a factor $(1-s_v)$ for a strange vector contributing to $r$ \cite{Seiden:1988rr}, and
a factor $(1-s_e)$ for a strange quark contributing to $e$.
The last term arises due to a combined mass/electromagnetic breaking of the flavour-$SU(3)$ symmetry.
This correction was first introduced by Isgur \cite{Isgur:1976hg} who analysed corrections to
$V\to P\gamma$ radiative decays through a parameter $x\equiv\mu_d/\mu_s$ which accounts for the expected difference in the $d$-quark and $s$-quark magnetic moments due to mass breaking.
The $V\to P\gamma$ amplitudes are precisely proportional to the electromagnetic contribution to the
$J/\psi\to VP$ decay, whose dominant decay occurs via $J/\psi\to\gamma\to VP$.
These results are reproduced in the $J/\psi\to VP$ amplitudes by means of the identification
$x\equiv 1-s_e$.
The general parametrization of amplitudes for $J/\psi\to VP$ decays is written in Table \ref{tableA}.
\begin{table}
\begin{tabular}{cc}
\hline
\tablehead{1}{c}{b}{Process} &
\tablehead{1}{c}{b}{Amplitude} \\
\hline
$\rho\pi$ 						& $g+e$ \\
$K^{\ast +}K^-+\mbox{c.c.}$		& $g(1-s)+e(2-x)$ \\
$K^{\ast 0}\bar K^0+\mbox{c.c.}$ 	& $g(1-s)-e(1+x)$ \\
$\omega_q\eta$ 					& $(g+e)X_\eta+\sqrt{2}r g[\sqrt{2}X_\eta+(1-s_p)Y_\eta]$ \\
$\omega_q\eta^\prime$ 			& $(g+e)X_{\eta^\prime}
                                                                       +\sqrt{2}r g[\sqrt{2}X_{\eta^\prime}+(1-s_p)Y_{\eta^\prime}]$ \\
$\phi_s\eta$ 					& $[g(1-2s)-2 e x]Y_\eta+r g(1-s_v)[\sqrt{2}X_\eta+(1-s_p)Y_\eta]$ \\
$\phi_s\eta^\prime$ 				& $[g(1-2s)-2 e x]Y_{\eta^\prime}
                                                                       +r g(1-s_v)[\sqrt{2}X_{\eta^\prime}+(1-s_p)Y_{\eta^\prime}]$ \\
$\rho\eta$ 					& $3e X_\eta$ \\
$\rho\eta^\prime$ 				& $3e X_{\eta^\prime}$ \\
$\omega_q\pi^0$ 				& $3e$ \\
$\phi_s\pi^0$ 					& $0$ \\
\hline
\end{tabular}
\caption{
General parametrization of amplitudes for $J/\psi\to VP$ decays.}
\label{tableA}
\end{table}
In order to obtain the physical amplitudes for processes involving $\omega$ or $\phi$ one has to incorporate corrections due to non-ideal $\omega$-$\phi$ mixing (see Notation).
Given the large number of parameters to be fitted, 13 in the most general case for 11 observables
(indeed 10 because there is only an upper limit for $\phi\pi^0$),
we perform the following simplifications.
First, we set the $SU(3)$-breaking corrections $s_v$ and $s_p$ equal to $s$.
This is motivated by the fact that all of them are due to the quark mass difference,
$m_{u,d}\neq m_s$, and thus expected to be of the same size.
Second, we fix the parameters $x=m_{u,d}/m_s$ and the vector mixing angle $\phi_V$ to the values obtained from a recent fit to the most precise data on $V\to P\gamma$ decays \cite{Escribano:2007cd},
that is $m_s/m_{u,d}=1.24\pm 0.07$ which implies $x=0.81\pm 0.05$ and $\phi_V=(3.2\pm 0.1)^\circ$.
The value for $x$ is within the range of values used in the literature, $0.62$ \cite{Baltrusaitis:1984fx},
$0.64$ \cite{Jousset:1988ni}, $0.70$ \cite{Morisita:1990cg}, and
the $SU(3)$-symmetry limit $x=1$ \cite{Coffman:1988ve}.
The value for $\phi_V$ is in perfect agreement (magnitude and sign)
with the value $\phi_V=(3.4\pm 0.2)^\circ$ obtained from the ratio
$\Gamma(\phi\to\pi^0\gamma)/\Gamma(\omega\to\pi^0\gamma)$ and the $\omega$-$\phi$ interference in $e^+e^-\to\pi^+\pi^-\pi^0$ data \cite{Dolinsky:1991vq}.
It is also compatible with the value $\phi_V^{\rm quad}=+3.4^\circ$
coming from the squared Gell-Mann--Okubo mass formula (see Ref.~\cite{Yao:2006px}).
Finally, we do not allow for gluonium in the $\eta$ and $\eta^\prime$ wave functions,
thus the mixing pattern of these two mesons is given only by the mixing angle $\phi_P$ (see Notation).
The case of accepting the presence of gluonium in the wave functions is still under study\footnote{
Work in preparation.}.

\section{Results}
We proceed to present the results of the fits.
We will also compare them with others results reported in the literature and in particular with the ones obtained in 1997 \cite{Bramon:1997mf}.
To describe data without considering the contribution from the doubly disconnected diagram (terms proportional to $rg$) has been shown to be unfeasible \cite{Coffman:1988ve}.
Therefore, it is required to take into account nonet-symmetry-breaking effects.
We have also tested that it is not possible to get a reasonable fit setting the $SU(3)$-breaking corrections to their symmetric values, i.e.~$s=0$ and $x=1$.
In addition, the value of $x$ is weakly constrained by the fit.
For that reason, we start fitting the data with $x=0.81$ (see above) and leave $s$ free.
The vector mixing angle $\phi_V$ is for the time being also set to zero.
The result of the fit gives $\phi_P=(35.9\pm 1.8)^\circ$ ---or $\theta_P=(-18.8\pm 1.8)^\circ$---
with $\chi^2/\mbox{d.o.f.}=17.4/4$.
In some analyses, the $SU(3)$-breaking contributions $s_v$ and $s_p$ are set to zero since they always appear multiplying $r$ and hence the products $r s_v$ and $r s_p$ are considered as second order corrections which are assumed to be negligible.
In this case, our fit gives $\phi_P=(35.6\pm 1.7)^\circ$ ---or $\theta_P=(-19.1\pm 1.7)^\circ$---
with $\chi^2/\mbox{d.o.f.}=20.2/4$, in agreement with
$\theta_P=(-19.1\pm 1.4)^\circ$ \cite{Jousset:1988ni},
$\theta_P=(-19.2\pm 1.4)^\circ$ \cite{Coffman:1988ve}, and
$\theta_P\simeq -20^\circ$ \cite{Morisita:1990cg}.
However, none of the former analyses include the effects of a vector mixing angle different from zero.
It was already noticed in Ref.~\cite{Bramon:1997mf} that these effects,
which were considered there for the first time, turn out to be crucial to find a less negative value of the
$\eta$-$\eta^\prime$ mixing angle.
If we take now the fitted value $\phi_V=+3.2^\circ$ (see above), one gets
$\phi_P=(37.8\pm 1.7)^\circ$ ---or $\theta_P=(-16.9\pm 1.7)^\circ$--- with $\chi^2/\mbox{d.o.f.}=17.4/4$
and
$\phi_P=(37.6\pm 1.6)^\circ$ ---or $\theta_P=(-17.1\pm 1.6)^\circ$--- with $\chi^2/\mbox{d.o.f.}=19.9/4$
for $s_v=s_p=s$ and $s_v=s_p=0$, respectively.
We have also performed a fit with $s_v=s$ and $s_p=0$ in order to have a fair comparison with
Ref.~\cite{Bramon:1997mf}.
In this particular case, one gets
$\phi_P=(38.5\pm 1.6)^\circ$ ---or $\theta_P=(-16.3\pm 1.6)^\circ$--- with $\chi^2/\mbox{d.o.f.}=18.5/4$,
in agreement with $\theta_P=(-16.9\pm 1.7)^\circ$ \cite{Bramon:1997mf}.
These new fits seem to confirm the strong correlation between the two mixing angles.
Once the $\omega$-$\phi$ mixing angle effects are taken into account the $\eta$-$\eta^\prime$ mixing angle becomes less negative.
The main drawback of the present analysis is the poor quality of the fits.
Investigating the contributions of each process to the $\chi^2$ one immediately realizes that the
$\rho\pi$ process contributes the most ---around 10 units in all the cases.
For instance, if $B(\rho\pi)=(16.9\pm 1.5)\%$ \cite{Yao:2006px} is replaced by its old value
$(12.8\pm 1.0)\%$ \cite{Barnett:1996hr} one gets for our best fit  $\theta_P=(-16.9\pm 1.6)^\circ$
with $\chi^2/\mbox{d.o.f.}=7.6/4$, which becomes an acceptable fit without changing the mixing angle.
In Table \ref{tableB}, we display the results of our best fit with $\phi_V=(3.2\pm 0.1)^\circ$.
As mentioned, the $\rho\pi$ channel causes the main disagreement between predicted and experimental data.
The fitted values of the parameters are also shown for completeness.

\section{Conclusions}
In summary,
we have performed an updated phenomenological analysis of $J/\psi\to VP$ decays
with the purpose of determining the quark content of the $\eta$ and $\eta^\prime$ mesons.
The conclusions are the following.
First, assuming the absence of gluonium in the $\eta$ and $\eta^\prime$ wave functions
the $\eta$-$\eta^\prime$ mixing angle is found to be
$\phi_P=(37.8\pm 1.7)^\circ$ in the quark-flavour basis or
$\theta_P=(-16.9\pm 1.7)^\circ$ in the octet-singlet basis.
This value is in agreement with very recent experimental measurements
\cite{Ambrosino:2006gk} and phenomenological estimates \cite{Thomas:2007uy}.
Second, the inclusion of vector mixing angle effects, not included in previous analyses, turns out to be crucial to get a less negative value for $\theta_P$.
This confirms our findings in Ref.~\cite{Bramon:1997mf}.
Finally, it is worth noticing that the recent reported values of $B(J/\psi\to\rho\pi)$
by the BABAR \cite{Aubert:2004kj} and BES \cite{Bai:2004jn} Collab.~seem to prevent us from obtaining a reasonable description of data.
Corroboration of these measurements would be highly desirable.


\begin{theacknowledgments}
I would like to express my gratitude to the SCADRON 70 Organizing 
Committee for the opportunity of presenting this contribution, and for the 
pleasant and interesting workshop we have enjoyed.
This work was supported in part by the Ramon y Cajal program,
the Ministerio de Educaci\'on y Ciencia under grant FPA2005-02211,
the EU Contract No.~MRTN-CT-2006-035482, ``FLAVIAnet'',
the Spanish Consolider-Ingenio 2010 Programme CPAN (CSD2007-00042), and
the Generalitat de Catalunya under grant 2005-SGR-00994.
\end{theacknowledgments}

\end{document}